\documentclass[10pt,twocolumn,aps,pre,superscriptaddress]{revtex4-2}

\usepackage[T1]{fontenc}
\usepackage{epsf,graphicx}
\usepackage{natbib}
\usepackage{multirow}
\usepackage{amsmath,amssymb}
\usepackage[bottom]{footmisc}
\usepackage{lipsum}
\usepackage{graphicx}
\usepackage{dcolumn}
\usepackage{upgreek}

\usepackage{bm}
\usepackage{flushend}
\usepackage{color}
\usepackage{xcolor}
\usepackage{esvect}
\usepackage{physics}
\usepackage{tcolorbox}

\usepackage{hyperref}
\hypersetup{
colorlinks=true, 
citecolor=magenta,
breaklinks=true, 
urlcolor= blue, 
linkcolor= blue, 
bookmarksopen=true, 
}

\begin{document}

\title{Diffraction of deep-water solitons}

\author{Filip Novkoski}
\email{filip.novkoski@fau.de}
\affiliation{Universit\'e Paris Cit\'e, CNRS,  Matière et systèmes complexes, F-75013 Paris, France}
\affiliation{PULS, Institute for Theoretical Physics, FAU Erlangen-Nürnberg, 91058, Erlangen, Germany}
\author{Loïc Fache}
\email{loic-joseph.fache@u-pariscite.fr}
\affiliation{Universit\'e Paris Cit\'e, CNRS,  Matière et systèmes complexes, F-75013 Paris, France}
\affiliation{Univ. Lille, CNRS, UMR 8523 - PhLAM - Physique des Lasers Atomes et Mol\'ecules, F-59 000 Lille, France}
\author{Félicien Bonnefoy}
\affiliation{Nantes Universit\'e, \'Ecole Centrale Nantes, CNRS, LHEEA, UMR 6598, F-44 000 Nantes, France}
\author{Guillaume Ducrozet}
\affiliation{Nantes Universit\'e, \'Ecole Centrale Nantes, CNRS, LHEEA, UMR 6598, F-44 000 Nantes, France}
\author{Jason Barckicke}
\affiliation{Universit\'e Paris Cit\'e, CNRS,  Matière et systèmes complexes, F-75013 Paris, France}
\author{François Copie}
\affiliation{Univ. Lille, CNRS, UMR 8523 - PhLAM - Physique des Lasers Atomes et Mol\'ecules, F-59 000 Lille, France}
\author{Pierre Suret}
\affiliation{Univ. Lille, CNRS, UMR 8523 - PhLAM - Physique des Lasers Atomes et Mol\'ecules, F-59 000 Lille, France}
\author{Eric Falcon}
\affiliation{Universit\'e Paris Cit\'e, CNRS,  Matière et systèmes complexes, F-75013 Paris, France}
\author{Stéphane Randoux}
\affiliation{Univ. Lille, CNRS, UMR 8523 - PhLAM - Physique des Lasers Atomes et Mol\'ecules, F-59 000 Lille, France}

\begin{abstract}
  Solitons are localized nonlinear wave packets that propagate without spreading because nonlinearity balances dispersion. Their robustness is well understood in effectively one-dimensional systems, but introducing additional spatial dimensions is generally expected to destabilize them or destroy their coherent character. Here we experimentally investigate how deep-water gravity-wave solitons behave when a controlled transverse degree of freedom is introduced through diffraction. Using a large-scale water-wave facility, we generate solitonic wave packets whose transverse structure is imposed across a segmented wavemaker through either a sharp slit or a smooth Gaussian apodization. The resulting two-dimensional wave fields are measured with high spatial resolution. Diffraction reshapes the transverse profile of the wave packet while its longitudinal dynamics retain the characteristic features of a soliton. Nonlinear spectral analysis confirms that the solitonic content is preserved along the direction of propagation, whereas the transverse evolution follows the linear Fresnel laws of diffraction. These observations reveal an unexpected coexistence of nonlinear soliton dynamics and classical wave diffraction.
\end{abstract}

\maketitle

Diffraction is a fundamental and universal consequence of wave propagation. Its quantitative description was instrumental in establishing the wave nature of light, most notably through Fresnel's formulation of wave theory based on Huygens’ principle~\cite{Fresnel1821}. In this framework, diffraction arises whenever a wave encounters spatial variations in impedance or boundary conditions, independent of the specific physical system. Accordingly, diffraction is observed not only in optics~\cite{Born-Wolf} but also in other wave systems, e.g., acoustic waves~\cite{Bekefi1953}, elastic waves~\cite{Guz1978}, and surface water waves~\cite{Stamnes2017}.

On the surface of water, diffraction plays a central role in coastal hydrodynamics, particularly in the interaction of sea waves with engineered structures such as breakwaters~\cite{Dalrymple1990}, which protect shorelines. The effects of edges, apertures, and gaps have long been investigated  theoretically~\cite{Penney1952,Dalrymple1988,Dalrymple1990,Yoshimi1978,Buccino2025} and experimentally~\cite{Blue1949,Pos1987}, but predominantly within the linear regime. In realistic ocean conditions, however, nonlinear effects become significant, and coherent structures such as envelope solitons and breathers may form~\cite{dysthe2008oceanic,onorato2021observation}, serving as models for the formation of rogue waves~\cite{Onorato2013,kharif2008rogue}. 

Solitons are coherent wave packets that are robust under essentially one-dimensional propagation. However, they are inherently sensitive to transverse perturbations—an unavoidable feature of natural environments—so their persistence in two dimensions is generally not expected~\cite{zakharov1973instability,ghidaglia1996nonexistence,kivshar2000self,pelinovsky2001mysterious, Deconinck2006,sulem2007nonlinear,Ablowitz2021}. Recent optical experiments have demonstrated that nonlinear wave packets may retain robust soliton dynamics in certain two-dimensional configurations~\cite{dieli2026observation}. How coherent water-wave solitons respond to the introduction of a transverse spatial degree of freedom, however, remains largely unexplored experimentally. 

The competition between nonlinear self-stabilization and transverse spreading raises a fundamental question: how does a soliton behave when it undergoes diffraction in a genuinely two-dimensional setting? In deep water, diffraction, dispersion, and nonlinearity act simultaneously, and the inclusion of transverse dynamics fundamentally modifies wave evolution. Whether a soliton loses its coherence or retains a recognizable structure under such conditions remains unresolved.

In this paper, we investigate the diffraction of deep-water solitons in two dimensions. We demonstrate experimentally that, despite transverse dynamics and the breaking of one-dimensional constraints, solitons closely follow the predictions of linear Fresnel diffraction while preserving their solitonic spectral content. By combining controlled laboratory experiments with numerical simulations, we quantify the transverse deformation of the wavefront and uncover an unexpected correspondence between linear diffraction laws and nonlinear coherent wave dynamics.

\section{Theoretical background}\label{sec:nls}
The quantity of interest in our study, which we measure experimentally, is the surface elevation of water $\eta(x,y,t)$. 
In the linear regime, two-dimensional surface waves with harmonic time dependence reads $\eta(x,y,t) = u(x,y)e^{i\omega t}$, where $\omega$ is the angular frequency and satisfies the Helmholtz equation for the spatial envelope $u(x,y)$ as~\cite{Stamnes2017,stamnes1981focusing}
\begin{align}\label{Helmholtz}
    \left(\partial_{xx}+\partial_{yy}+k^2\right)u(x,y)=0,
\end{align}
with $k=2\pi/\lambda$ the wavenumber.

\begin{figure}[t!]
\centering
\includegraphics[width=\linewidth]{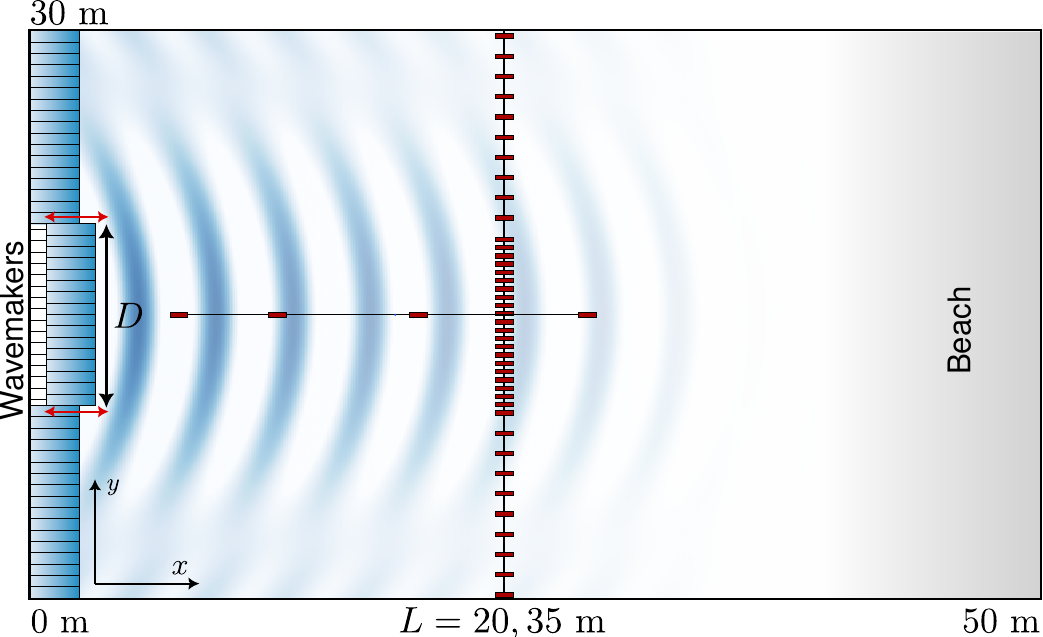}\vspace*{0.3em}
\includegraphics[width=\linewidth]{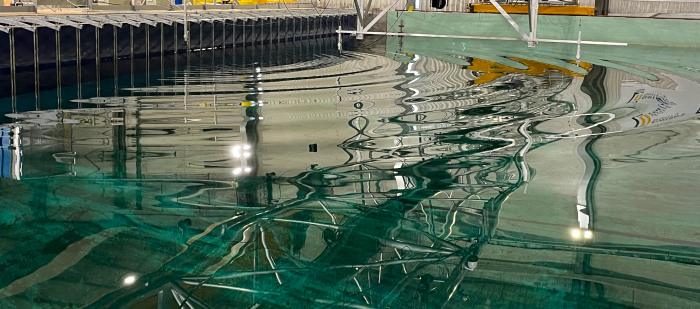}
\caption{Experimental set-up. Top: Schematic representation (not to scale) of the $3$D water tank (top view) used in the experiments. The transverse profile of the generated waves can be carefully shaped using 48 computer-controlled segmented wavemakers placed along the y axis, at $x=$~0 m.  Horizontal red bars: $45$ wave elevation probes are placed at discrete propagation distances $x = 1$–$25\,\mathrm{m}$ and transverse positions $y \in [0.1, 29.64]\,\mathrm{m}$ with a non-uniform spacing, as indicated by the probe array. The wavelength of the carrier wave is approximately $\lambda_c\simeq 1.3$ m. Bottom: Image of a diffracting wave with an opening of $D=1.2$ m (i.e. $2$ flaps out of $48$), see Movies S1 and S2.}
\label{fig:setup}
\end{figure}

The diffraction of a monochromatic wave by an aperture located at $x=0$ is described by the Fresnel–Kirchhoff integral, which in two dimensions yields the field $u(x,y)$ beyond the aperture as~\cite{stamnes1981focusing}
\begin{align}\label{eq:fresnel}
u(x,y)= \int_{-\infty}^{\infty}p(y_0)\,\left(-\frac{k x}{2 i\, r}\right)H_1^{(1)}\!\left(k r\right)\,\mathrm{d}y_0 ,
\end{align}
where $r=\sqrt{x^2+(y-y_0)^2}$ and $H_1^{(1)}$ is the first-order Hankel function of the first kind, and $y_{0}$ is the transverse coordinate along the aperture plane at $x=0$. The function $p(y_{0})$ represents the incident field distribution at the aperture, determined by the slit geometry and, in our experiments, will be imposed by the number of active wavemakers (see Fig.~\ref{fig:setup}). In the numerical evaluation of the integral, we adopt the Kirchhoff approximation, setting $p(y_{0}) = 1$ within the aperture and $p(y_{0}) = 0$ outside.

However, at higher wave amplitudes the linear description breaks down, as nonlinear effects substantially modify wave propagation. Under the assumption of paraxial propagation along the $x$-direction, the evolution of weakly nonlinear deep-water wave packets is instead governed by the (2D+1) hyperbolic nonlinear Schr\"{o}dinger equation (HNLSE)~\cite{zakharov1968stability,ghidaglia1993nonelliptic,Osborne2002}. This equation describes the dynamics of the complex envelope $A(x,y,t)$ of a carrier wave with wavenumber $k_0$ propagating along the $x$-direction with carrier frequency $\omega_0 = \sqrt{g k_0}$, where $g$ is the gravity acceleration.
\begin{align}\label{eq:nls_2d_full}
  \partial_x A+\frac{1}{c_g}\partial_t A=-\frac{i}{4k_0}\left(\partial_{xx}A-2\partial_{yy}A\right)-ik_0^3\vert A\vert^2 A,
\end{align}
where $c_g = \mathrm{d}\omega/\mathrm{d}k \big|_{k_0}$ is the group velocity. The surface elevation $\eta(x,y,t)$ can be approximated to the first order by~\cite{Osborne2002}
\begin{align}
    \eta(x,y,t) = \mathrm{Re}\!\left\{ A(x,y,t)\,
e^{i(k_0 x - \omega_0 t)} \right\}.
\end{align}
At leading order in the weakly nonlinear regime, the envelope propagates at the group velocity $c_g$, so that $\partial_t A + c_g \partial_x A \simeq 0$~\cite{Osborne2002,Chabchoub2019}. This relation is used to rewrite the longitudinal second-order dispersive term, while the transport operator $\partial_x + c_g^{-1}\partial_t$ is kept unchanged as it defines the envelope evolution. HNLSE thus reads
\begin{align}\label{eq:nls_2d}
  \partial_x A + \frac{1}{c_g} \partial_{t}A = -i\left(\frac{1}{g} \partial_{tt}A - \frac{1}{2k_0} \partial_{yy}A \right) - ik_0^3 |A|^2 A.
\end{align}
While the waves remain focusing in the longitudinal direction (as shown below), they are defocusing transversely, raising questions about their stability in the presence of this additional spatial dimension~\cite{zakharov1973instability,Deconinck2006}. To explore this regime and compare with experiments, we numerically integrate~\eqref{eq:nls_2d} (see Methods).

If the solution is assumed to be independent of the transverse direction $y$, we recover the (1D+1) focusing nonlinear Schr\"{o}dinger equation (NLSE), corresponding to purely unidirectional wave propagation along the $x$ axis, as
\begin{align}\label{eq:nls_1d}
  \partial_x A + \frac{1}{c_g} \partial_{t}A = -\frac{i}{g} \partial_{tt}A - ik_0^3 |A|^2 A,
\end{align}
which, unlike~\eqref{eq:nls_2d_full} and~\eqref{eq:nls_2d}, is integrable and can be solved through the use of the inverse scattering transform (IST)~\cite{novikov1984theory,ablowitz1973nonlinear,yang2010nonlinear}. The well-known fundamental single-soliton solution of~\eqref{eq:nls_1d} can be written as~\cite{zakharov1968stability,Remoissenet2013,cazaubiel2018coexistence},
\begin{align}\label{eq:soliton}
\eta(x,t) =& a \,
\operatorname{sech}\!\left[
\frac{a k_0 \omega_0}{\sqrt{2}}
\left(t - \frac{x}{c_g} \right)
\right]\times \nonumber\\
&\cos\!\left[
\omega_0 t + k_0
\left(
1 + \frac{k_0^2 a^2}{4}
\right)
x
\right],
\end{align}
where $a$ is the maximal soliton envelope amplitude. Within the IST framework, a soliton is characterized by a discrete complex eigenvalue $\zeta$ of the associated scattering problem. The imaginary part of this eigenvalue determines the soliton amplitude, while its real part is related to its propagation velocity. The eigenvalue is obtained by solving the corresponding scattering problem (see Methods).

\begin{figure*}[t!]
\centering
\includegraphics[width=\linewidth]{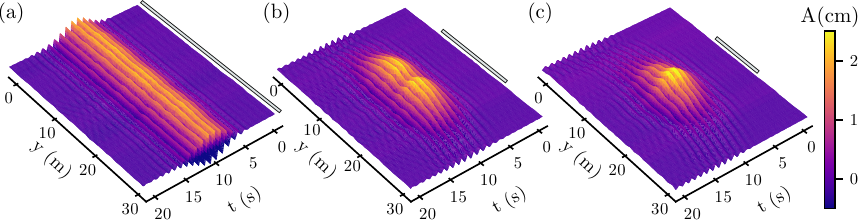}
\caption{\label{fig:3d}Measured surface elevation $\eta(y,t)$ of a soliton of steepness $\epsilon=k_0a=.097$ measured by the wave probes at $L=20$~m for three different slit openings (see white rectangles) $D=30$(a),15(b) and 10(c)~m. For the largest opening, (a), we observe the classical profile of a 1D NLSE soliton [\eqref{eq:soliton}]. Decreasing the aperture width, as seen in (b) and (c) reduces the transverse ($y$) size of the soliton, however we still observe a coherent structure in the basin center. Additionally, we observe the appearance of distinct minima and maxima in the transverse profile, consistent with classical wave diffraction. }
\end{figure*}


\section{Experimental setup}\label{sec:exp}
Experiments were performed in the large-scale wave basin ($50$~m long $\times$ $30$~m wide $\times$ $5$~m deep) of Ecole Centrale de Nantes, France. The experimental setup is sketched in Fig.~\ref{fig:setup}. The wave generation mechanism consists of $48$ independently controlled wavemakers (flaps of width 0.62~m, hinged 2.8 m from the free surface) located at one end of the basin, i.e. at $x=0$. An absorbing, sloping beach is located at the opposite end.

We focus on two types of soliton waveforms, either a slit-diffracting soliton or a Gaussian beam (a soliton with a transverse Gaussian profile along the $y$-direction, see Fig.~\ref{fig:setup}). Here, the term "slit" will be used in analogy with optical diffraction, as the aperture is implemented by selectively driving a finite set of neighbouring wavemakers. To generate a diffracting soliton, the wavemakers are driven by a monochromatic carrier of fixed frequency $f_0=1.1$~Hz [i.e., a fixed carrier wavelength of $\lambda_0 = 2\pi/k_0 = g / (2\pi f_0^2)\simeq 1.3$~m], amplitude-modulated by a hyperbolic secant following the 1D NLSE solitonic solution of~\eqref{eq:soliton} at $x=0$. The carrier wavenumber $k_0$ is kept fixed, while the soliton maximal envelope amplitude $a$ is varied around typical values of $0.4-2\,\mathrm{cm}$ and has a typical size $L_x = gT_0/\omega_0\in [5.38,25]~\mathrm{m}$, where $T_0$ is the typical duration of the soliton. The corresponding carrier steepness $\epsilon \equiv k_0 a$ is explored over the range $[0.019,\,0.044,\,0.070,\,0.097,\,0.127]$, allowing us to probe regimes from weak to strong nonlinearity. If all wavemakers are driven in phase, a 1D-NLSE soliton with a transverse extension covering the whole width (30 m) of the water tank is generated and propagates towards the beach. By altering the number of working wavemakers, we can change the diffraction aperture, $D$ (see top of Fig.~\ref{fig:setup}), to observe its impact on the soliton propagation and its diffraction. The parameter $D$ is varied from a small $0.6$~m aperture to the full $30$~m width of the basin. 

To generate a Gaussian beam, the carrier wave is now amplitude-modulated along the wavemakers, i.e. in the transverse $y$-direction. By weighting the driving amplitudes of the different wavemakers, we implement a Gaussian apodization of the initial soliton transverse profile, enabling a controlled smoothing of the aperture edges.

\begin{figure*}[!t]
    \centering
    \includegraphics[width=\textwidth]{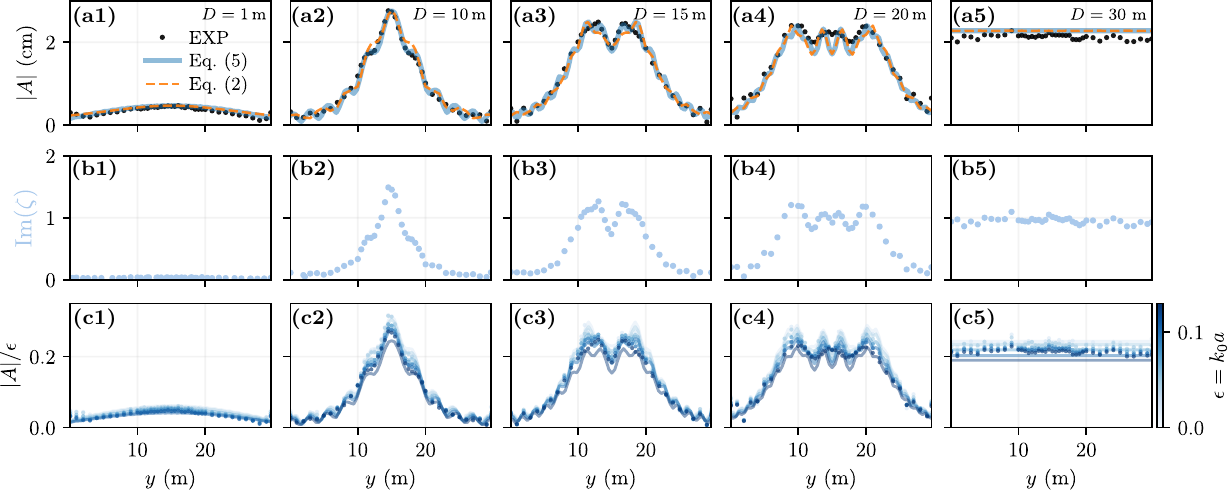}
    \caption{\label{fig:comparison}
    Transverse structure of diffracting deep-water solitons. Amplitude and IST profiles with comparison with HNLSE and Helmholtz diffraction. The columns correspond to different aperture widths $D=1,\,10,\,15,\,20,$ and $30~\mathrm{m}$ (from left to right), measured at a fixed propagation distance $L=20~\mathrm{m}$.  
    (a1--a5) Transverse envelope amplitude $|A(y)|$. Black dots: experimental data ($\epsilon = 0.097)$; solid blue curves: numerical simulations of the HNLSE~\eqref{eq:nls_2d}; dashed orange curves~\eqref{eq:fresnel}: linear Helmholtz diffraction by a rectangular slit of width $D$ evaluated for the carrier wavenumber $k_0$. The Helmholtz predictions are rescaled by a single multiplicative factor to match the experimental peak amplitude in each panel. 
    (b1--b5) IST spectra extracted from the longitudinal wave field and represented as function of the transverse coordinate $y$. Blue points correspond to imaginary part of the discrete eigenvalue, $\mathrm{Im}(\zeta)$, which characterizes the soliton amplitude. The presence of discrete eigenvalue across $y$ indicate the solitonic nature of the wave packet.
    (c1--c5) Steepness-normalized transverse amplitude $|A(y)|/\epsilon$ for different soliton amplitudes, $a\in[0.37,2.61]\, \mathrm{cm}$, generated with the same aperture $D$. For each run, the experimental curves (markers) and the associated HNLSE profiles (solid curves) are shown for several steepness values $\epsilon=k_0 a \in [0.02,0.13]$ (the color encodes the soliton steepness, as indicated by the colorbar on the right). 
    }
\end{figure*}

The surface elevation $\eta(t)$ is recorded using an array of $45$ resistive wave probes. Of these, 41 are positioned along a straight transverse line located at a selectable distance of either $L$=20~m or $L$=35~m from the wavemakers. The probes are spaced 1~m apart, except for the central 23 probes, which are separated by 0.5~m to enhance spatial resolution. The arrangement is shown at the top of Fig.~\ref{fig:setup}. Four additional probes are located in the main propagation direction. The sensors provide a vertical resolution of 0.1~mm, a frequency bandwidth of 20~Hz, and are sampled at 128~Hz.
In order to look for the influence of dispersion, diffraction and nonlinearity in the present configuration, it is useful to introduce the associated characteristic lengths and their range of accessible values in our experiment. The dispersive term in~\eqref{eq:nls_2d} defines the dispersive length $L_\text{disp} = g T_0^2 \in [141,3285]~\mathrm{m}$. The transverse diffraction term yields a diffraction length $L_\text{diff} = 2k_0D^2\in [9.7,3895]~\mathrm{m}$. Finally the nonlinear term defines a nonlinear length $L_\text{NL} = 1/ k_0^3 a^2 \in [12.7,569]~\mathrm{m}$.

\section{Slit diffraction of solitons}

As described above, we generate wave packets whose longitudinal dynamics correspond to exact single-soliton solutions~\eqref{eq:soliton} of the (1D+1) NLSE. By varying the aperture width $D$, we introduce controlled transverse diffraction. Figure~\ref{fig:3d} shows the resulting wave field for three different aperture openings. For the maximal opening (Fig.~\ref{fig:3d}a), when all wavemakers are active, the classical NLSE single soliton described by~\eqref{eq:soliton} is recovered. As the aperture width is reduced (Fig.~\ref{fig:3d}b–c), the wave packet remains localized while becoming transversely confined to the central region of the basin, over a width consistent with the imposed opening. This is complemented by a now very visible curved wavefront. Additionally, we can notice the emergence of clear maxima and minima in the soliton transverse $y$-profile, demonstrating the appearance of a diffraction pattern.

These features are more clearly observed in Fig.~\ref{fig:comparison}(a1--a5), where the transverse envelope amplitude $A(L,y)$ at a distance $L=20$~m broadens as the aperture width $D$ increases, while progressively developing the characteristic diffraction pattern with alternating transverse maxima and minima. The measured amplitude profiles (dots) are compared with numerical solutions of~\eqref{eq:nls_2d} (solid lines), showing excellent agreement and confirming that the HNLSE accurately captures soliton diffraction for varying apertures. We further compare the data with the classical Fresnel-Kirchhoff prediction for linear monochromatic waves given by~\eqref{eq:fresnel} (dashed orange lines) derived from the Helmholtz equation. Remarkably, although the soliton is an intrinsically nonlinear wave packet, its diffraction from a slit is quantitatively described by the classical linear theory.

To further elucidate the soliton behavior under slit diffraction, the longitudinal measurements are analyzed using the inverse scattering transform (IST), as detailed in Methods. This approach provides a nonlinear spectral characterization of the (1D+1) NLSE along the longitudinal direction at each transverse position. Unlike envelope measurements alone, the IST yields direct access to the discrete eigenvalues that quantify solitonic content.

The resulting transverse dependence of the discrete complex eigenvalue $\zeta(y)$ is shown in Fig.~\ref{fig:comparison}(b1--b5) for the soliton amplitude, i.e. $\text{Im} [\zeta(y)]$. For sufficiently large aperture widths $D$, well-defined eigenvalues persist across the transverse direction, even as the envelope exhibits pronounced diffraction. This indicates that while the finite aperture reshapes the transverse structure of the wavefront, the longitudinal dynamics at each transverse position remain governed by soliton behavior.

\begin{figure}[t!]
    \centering
    \includegraphics[width=\linewidth]{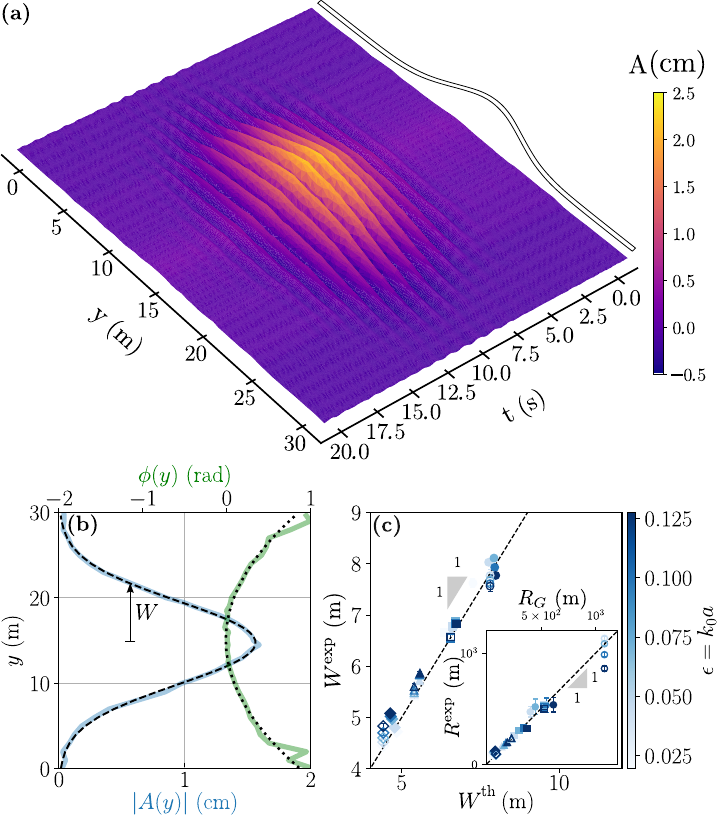}
    \caption{(a) Measured surface elevation $\eta(y,t)$ of a solitonic Gaussian beam with $\epsilon \approx 0.072$ and waist $W_0\approx 6.43$ m. The beam is well localized in the transverse direction, and unlike the slit-diffracted soliton, we observe a much more regular wavefront due to the Gaussian apodization (white bar). (b) Transverse cut of a reference soliton at $x=20~\mathrm{m}$, showing the measured envelope amplitude $|A(y)|$ (blue, solid) together with a Gaussian fit (black, dashed), and the corresponding unwrapped transverse phase profile $\phi(y)$ (green, solid) with a parabolic fit (black, dotted), from which the wavefront radius of curvature is extracted. The arrow indicates the fitted transverse waist $W$ of the Gaussian envelope. (c) Measured waist $W^{\mathrm{exp}}$ as a function of the theoretical prediction $W^{\mathrm{th}}$ (\eqref{eq:gaussbeam}) for different initial waists $W_0$.  Inset : Measured radius of curvature $R^{\mathrm{exp}}$ versus the theoretical radius $R_{\mathrm{G}}$ (\eqref{eq:gaussbeam}). In panel (c) and its inset, filled symbols correspond to experiments and open symbols to HNLSE simulations. Marker shapes encode different initial waists ($W_0\approx \diamond 3.87,\, \triangle 5.15,\, \square 6.43,$ and $\circ7.73~\mathrm{m}$), while the color scale on the right indicates the soliton steepness $\epsilon = k_0 a$.}
    \label{fig:3d_gauss}
\end{figure}

In contrast, for small aperture widths (see Fig.~\ref{fig:comparison}b1), no discrete eigenvalues are detected, demonstrating the absence of solitonic content. This reveals the existence of a threshold aperture size below which the wave field becomes purely dispersive and soliton dynamics are lost.

These observations demonstrate that transverse diffraction induced by a finite aperture reshapes the soliton front without destroying its longitudinal soliton character. We therefore observe the coexistence of one-dimensional integrable soliton dynamics along the propagation direction $x$ with transverse linear spreading. The origin of this behavior can be understood physically. In a purely dispersive medium, a wave packet would broaden along the propagation direction, and diffraction at an aperture would generate the superposition of Fresnel patterns associated with its spectral components. Here, however, the IST analysis shows that the wave packet retains its solitonic content: dispersion along $x$ is continuously balanced by nonlinearity, so that the packet remains effectively non-dispersive in the longitudinal direction. Transversely, by contrast, the governing equation is defocusing and does not provide comparable nonlinear self-confinement. The lateral evolution is therefore dominated by diffraction. Because the longitudinal structure remains intact, the transverse field behaves as that of a coherent, non-dispersive object and is accurately described by classical Fresnel diffraction theory derived from the Helmholtz equation. Consequently, the soliton undergoes essentially linear transverse diffraction while preserving its nonlinear identity along the direction of propagation.

Finally, we examine the influence of the soliton amplitude on the diffraction pattern, as shown in Fig.~\ref{fig:comparison}(c1--c5). The amplitudes $|A(y)|$ are normalized by the wave steepness $\epsilon$ and compared with numerical simulations of~\eqref{eq:nls_2d}. The resulting diffraction patterns exhibit clear self-similarity across different probed nonlinearities (see right colorbar) and aperture widths $D$. This invariance indicates that, for sufficiently large $D$, the longitudinal soliton dynamics remain intact, with transverse diffraction largely independent of the soliton amplitude.

\section{Solitons with a transverse Gaussian profile\label{sec:gauss}}

In the previous section, diffraction was induced by imposing a sharp transverse truncation of the wavemaker motion, corresponding to the slit geometry. We now move beyond this configuration by importing concepts from Gaussian beam optics to probe soliton diffraction from a different perspective. In optics, Gaussian beams provide a fundamental and analytically tractable description of diffraction and beam spreading~\cite{svelto2010principles}. We reproduce this geometry experimentally on the surface of deep water by imposing a Gaussian apodization across the wavemaker array at $x=0$ and centered on the basin $y$-axis. The resulting wave field constitutes a genuine hydrodynamic analogue of an optical Gaussian beam, with a crucial distinction: its longitudinal dynamics remain governed by the NLSE and retain a solitonic character, while the transverse envelope is initially Gaussian with waist $W_0$. This construction allows us to test directly whether the well-established propagation laws of Gaussian beams extend to nonlinear soliton wave packets. An example of the measured elevation of a soliton generated this way is displayed in Fig.~\ref{fig:3d_gauss}a, showing a clear localization both in the transverse $y$-direction, as well as in time, i.e in the $x$-direction.

\begin{figure}[t!]
    \centering
    \includegraphics[width=\linewidth]{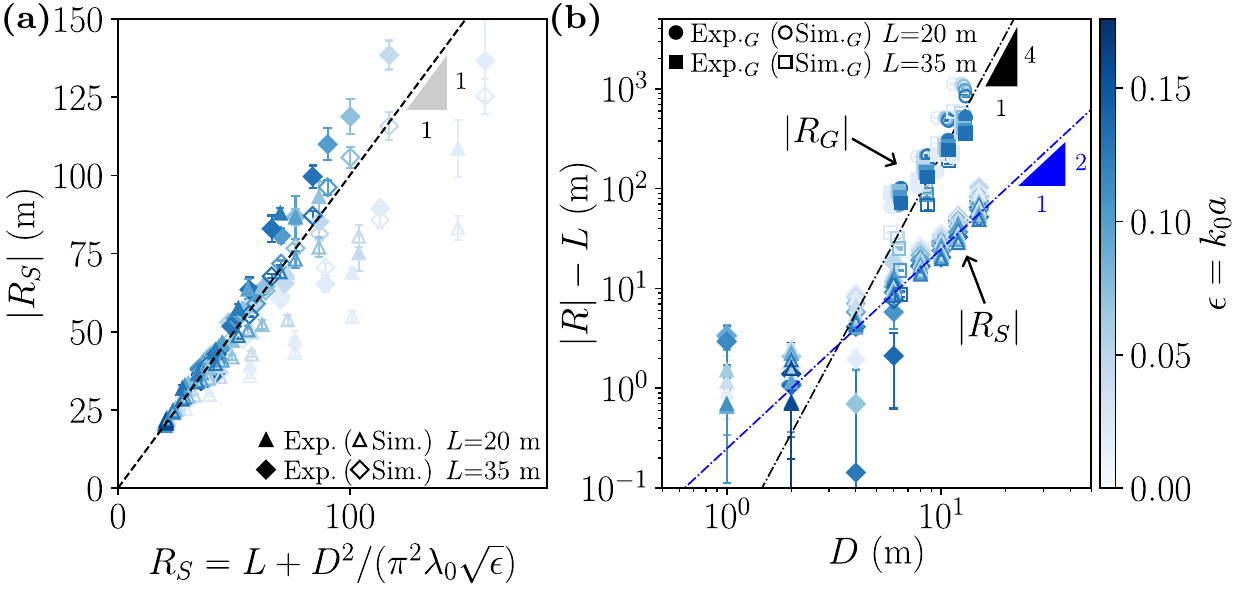}
    \caption{\label{fig:curvature}
(a) Absolute value of the soliton wavefront radius of curvature $|R_S|$, extracted from the carrier-wave front, as a function of the rescaled variable $L + D^2/(\pi^2 \lambda_0 \sqrt{\epsilon})$, where $L$ is the propagation distance, $D$ the transverse aperture (slit) width, $\lambda_0$ the carrier wavelength, and $\epsilon = k_0 a$ the steepness. Full (open) symbols denote experiments (HNLSE simulations). Triangles correspond to $L=20\,\mathrm{m}$ and diamonds to $L=35\,\mathrm{m}$. The color encodes $\epsilon$. Error bars indicate the standard deviation over all fits of the individual carrier-wave oscillations. The black dashed line (slope $1$) highlights the empirical collapse of~\eqref{eq:scaling}. (b) (log--log axes): Curvature $|R|-L$ versus $D$. Circles ($L=20\,\mathrm{m}$) and squares ($L=35\,\mathrm{m}$) show Gaussian-apodized beams ($R_G$) extracted from quadratic phase fits (full: experiments; open: HNLSE simulations), with $D= 2W_0 \sqrt{\ln 2}$ for five different steepness values. Black dash-dotted line correspond to~\eqref{eq:gaussbeam} with $L=35$~m and the blue one to the $D^2$ scaling expected from~\eqref{eq:scaling} with $\epsilon=0.1$. Error bars for these points originate from the uncertainties of the quadratic phase fits, as estimated from the covariance matrix of the fit parameters.}
\end{figure}

We expect that the evolution of the wavefield follows the standard paraxial Gaussian-beam propagation laws~\cite{svelto2010principles} 
\begin{align}
W(x)=W_0\sqrt{1+\left(\frac{x\lambda_0}{\pi W_0^2}\right)^2},
\
R_G(x)=x\left[1+\left(\frac{\pi W_0^2}{x\lambda_0}\right)^2\right],
\label{eq:gaussbeam}
\end{align}
where $W(x)$ is the waist at $x$ extracted from the soliton envelope $|A(x,y)|$, $R_G(x)$ is the radius of curvature inferred from the transverse phase, and $\lambda_0$ is the carrier wavelength. Similar propagation laws have been observed experimentally for the manipulation of water waves using electrostriction~\cite{mouet2023comprehensive}. 

We show a measured transverse profile of a Gaussian apodized soliton in Fig.~\ref{fig:3d_gauss}b, demonstrating that it indeed retains extremely well a Gaussian form after propagation, both in its amplitude and its phase. The curvature of the wavefront is directly extracted from the transverse phase of the carrier wave, which is locally well described by a quadratic fit $\phi(y) \simeq \phi_0 + \alpha y^2$ (see dotted line), yielding a radius of curvature $R^{\text{exp}} \equiv k_0(2\alpha)^{-1}$. While such a characterization is standard in the context of Gaussian beams in optics, it is most often obtained indirectly, either from the evolution of the beam waist under the Gaussian-beam assumption~\cite{svelto2010principles}, or via dedicated wavefront-sensing techniques such as Shack--Hartmann sensors or Talbot--effect methods~\cite{neal1996amplitude,zavalova2002shack,thul2020spatially,goloborodko2023wavefront,kotov2023talbot} rather than from a direct measurement of the carrier phase. In contrast, our hydrodynamic system provides direct access to the full spatial-temporal wavefield, allowing the curvature of the carrier-wave front to be measured. By repeating measurements for different widths as well as nonlinearities, we confirm in Fig.~\ref{fig:3d_gauss}c that such solitons obey perfectly the relationships of~\eqref{eq:gaussbeam}, both experimentally (full symbols) and numerically (open symbols), giving the first empirical deep-water realization and control of focused Gaussian beams.

\section{Curvature radius of a diffracted soliton}
Using the direct availability of the wavefront curvature, we use it to further study the slit-diffracted solitons and compare to Gaussian-apodized solitons. At fixed $L$, individual wavefront oscillations are locally well captured by a quadratic phase profile in $y$, from which a curvature radius can be defined (see the procedure illustrated for Gaussian-apodized solitons in Fig.~\ref{fig:3d_gauss}b).

Figure~\ref{fig:curvature}a shows all curvature measurements for slit-diffracted solitons. Each dataset corresponds to a given propagation distance ($L=20$ or $35$~m), aperture width $D$, and steepness $\epsilon$, with colors encoding $\epsilon$ and open symbols denoting HNLSE simulations. Remarkably, despite such a wide range of parameters, all points align remarkably well onto a single straight line of slope $1$, (black dashed line) once the abscissa is rescaled according to 
\begin{align}\label{eq:scaling}
   R_S = L + \frac{D^2}{\pi^2\lambda\sqrt{\epsilon}},
\end{align}
which, to the best of our knowledge, is a so far unreported relationship. This collapse shows first that the effect of aperture enters predominantly through a quadratic dependence, $R_S - L \propto D^2$. Second, it reveals a strong influence of nonlinearity through the factor $1/\sqrt{\epsilon}$, i.e., at fixed $L$ and $D$, increasing $\epsilon$ decreases $R_S$, meaning that the wavefront becomes more curved. This trend is consistent with an effective nonlinear focusing of the diffracting wavefront~\cite{zakharov1968instability,akhmanov1968self}. In other words, while the transverse spreading remains governed by the finite aperture, nonlinear effects renormalize the local wavefront geometry. 

Figure~\ref{fig:curvature}b places this slit-diffracted behaviour (diamond and triangle symbols) in perspective by comparing it to Gaussian-apodized solitons (square and circle symbols, see Sec.~\ref{sec:gauss}), where the initial transverse profile is smooth rather than truncated. In that case, as shown in Fig.~\ref{fig:3d_gauss}c both the waist and the curvature follow the standard Gaussian-beam propagation predictions~\eqref{eq:gaussbeam}, leading to a much stronger dependence on the transverse size, namely $R_G \propto D^4$ since $W_0 = D/(2\sqrt{\ln 2})$.  
The coexistence of a $D^2$ scaling for the slit-diffracted soliton $R_S$ and a $D^4$ scaling for the Gaussian-apodized soliton $R_G$ highlights that the initial transverse shaping at $x=0$ plays a decisive role in setting the wavefront geometry. 

At present, a theoretical description of the empirical scaling~\eqref{eq:scaling} remains an open question.

\section{Conclusion}

We have investigated how a one-dimensional deep-water soliton evolves when subjected to two-dimensional diffraction. Drawing inspiration from optics, where the transverse profile of a beam is commonly shaped using a slit or Gaussian apodization, we extended this idea to water waves by introducing a controlled transverse degree of freedom while maintaining the longitudinal coherent soliton structure near the wavemakers.

Despite the presence of the additional spatial dimension, the wave packet retains its solitonic character. Transverse spreading is remarkably well described by linear Fresnel diffraction theory, while nonlinear spectral signatures associated with soliton dynamics persist along the propagation direction, as demonstrated clearly by the IST method. Linear diffraction and nonlinear coherence therefore coexist within a single wave structure.

The robustness of coherent structures under dimensional extension is a central question across many domains, since higher-dimensional effects are known to trigger transverse instabilities~\cite{zakharov1973instability} or qualitatively modify soliton dynamics~\cite{kivshar2000self,moll2003self,sulem2007nonlinear}. This issue is of direct and applicable interest in nonlinear optics and Bose--Einstein condensates, as well as plasma physics and ocean wave dynamics. Our results demonstrate that the transition from integrable one-dimensional dynamics to genuinely two-dimensional behavior is rather progressive than abrupt. More broadly, this controlled experimental platform provides a quantitative framework to explore weakly non-integrable regimes and to test perturbative approaches that attempt to bridge ideal integrable models and realistic nonlinear wave systems~\cite{kaup1976perturbation,karpman1977perturbation,Malomed1989,gelash2024bi,Falsi2024}.

\section{Methods}
\subsection*{Integration of the 2D NLSE}\label{subsec:methods}
We integrate numerically the hyperbolic (2D+1) NLSE~(\eqref{eq:nls_2d}) written in the retarded-time frame, for the complex envelope $A(x,t,y)$ using a pseudo-spectral scheme in the $(t,y)$ plane. The field is discretized on a rectangular periodic domain $t\in[-L_t/2,L_t/2]$, $y\in[-L_y/2,L_y/2]$ with uniform grids of size $N_t\times N_y$; derivatives $\partial_{tt}$ and $\partial_{yy}$ are evaluated in Fourier space using fast Fourier transforms, while the cubic term $|A|^2A$ is computed in physical space. In the simulations we use, $L_t = 80~\mathrm{s}$, and $L_y = 110~\mathrm{m}$ with $N_t = 512$ and $N_y=256$ grid points. The resulting system of ordinary differential equations in the propagation variable $x$ is advanced with an explicit adaptive Runge--Kutta integrator (DOP853)~\cite{hairer1993solving}. The initial condition reproduces the experimental forcing as a soliton waveform, with a transverse profile (super-Gaussian, Gaussian or other) matching the experimental aperture. Parameters are set by the experimental carrier frequency ($k_0=\omega_0^2/g$) and steepness $\epsilon=k_0 a$.

\subsection*{The IST spectrum of experimental data}
For a solution of the nonlinear Schrödinger equation (NLSE) comprising N solitons, the discrete spectrum consists of N complex eigenvalues 
$\zeta_n$, each associated with a complex norming constant $C_n$ that characterizes the corresponding phase. Collectively, these quantities constitute the 
complete set of scattering data for the solution.

The discrete spectrum is determined by solving the eigenvalue problem associated with the Lax pair formulation of the NLSE. Each eigenvalue admits a direct physical interpretation: its real part corresponds to the soliton velocity, while its imaginary part determines the soliton amplitude.

In the focusing case of the NLSE, the associated eigenvalue problem reduces to the Zakharov–Shabat spectral problem~\cite{shabat1972exact},
\begin{align}
\hat{\mathcal{L}}\,\Phi = \zeta\,\Phi,
\qquad
\hat{\mathcal{L}} =
\begin{pmatrix}
 i \partial_\xi & - i \psi \\
- i \psi^{*}   & \;\, -i \partial_\xi
\end{pmatrix},
\label{eq:zs_problem}
\end{align}
where $\Phi(\xi, \zeta)$ is a vector wave function. $\zeta \in \mathbb{C}$ represent the eigenvalues composing the discrete spectrum associated with the soliton content of the field $\psi(\tau,\xi)$ that is measured at some given evolution time $\tau$ (or propagation distance $x$ in the experiment) and at each $y$-coordinate. Connection between the physical envelope $A(x,t)$ and the dimensionless variables used in the problem of~\eqref{eq:zs_problem} is given by, $\tau = x k_0^3 a^2 / 8$, $\xi = \sqrt{k_0^2 \omega_0^2 a^2/8}\left( t- \frac{x}{c_g} \right)$ and $\psi(\tau,\xi) = A(x,t)/(a/2)$, see~\cite{Suret2020,Fache2024}.

We reconstruct the slowly varying complex envelope $A(x;y)$ from the surface elevation $\eta (x,t;y)$ by demodulating the carrier wave at frequency $\omega_0$ and wavenumber $k_0$ through the use of the Hilbert transform, yielding both the amplitude and phase of the envelope. The envelope is then rescaled using the standard deep-water NLSE normalization given above. For each transverse position $y$, the reconstructed envelope $A(x;y)$ is treated as an initial condition of the Zakharov-Shabat spectral problem~\eqref{eq:zs_problem}. The Zakharov-Shabat problem is then solved numerically for each transverse position $y$ using the Fourier collocation method, following the procedure described in~\cite{yang2010nonlinear} and previously successfully used experimentally in~\cite{Suret2020,Fache2024}.

\begin{acknowledgments}
We thank A. Levesque, S. Mazo, B. Pettinotti (ECN) for their technical help on the experimental setup. This work was partially supported by the Agence Nationale de la Recherche through the SOGOOD (Grant No. ANR-21-CE30-0061) project, the Simons Foundation MPS-WT No. 651463 project. F. C., P. S. and S. R. acknowledge the support of the CDP C2EMPI, as well as the French State under the France-2030 programme, the University of Lille, the Initiative of Excellence of the University of Lille, the European Metropolis of Lille for their funding and support of the R-CDP-24-004-C2EMPI project. F.N thanks the Humboldt foundation for a postdoctoral fellowship.
\end{acknowledgments}

\bibliographystyle{apsrev4-1}
\bibliography{biblio}

\end{document}